\documentstyle[eqsecnum,aps,preprint,psfig]{revtex}
\begin{document}
\tightenlines
\draft
\title{Relativistic Brueckner-Hartree-Fock calculations with explicit
intermediate negative energy states}

\draft

\author{F. de Jong and H. Lenske}
\address{Institut f\"ur Theoretische Physik, Universit\"at Giessen,
35392 Giessen, Germany}
\date{\today}

\maketitle
\begin{abstract}
In a relativistic Brueckner-Hartree-Fock calculation we include 
explicit negative-energy states in the two-body propagator. 
This is achieved by using the Gross spectator-equation,  
modified by medium effects.
Qualitatively our results compare well with other RBHF calculations.
In some details significant differences occur, e.g, our equation of 
state is stiffer and the momentum dependence of the self-energy components
is stronger than found in a reference calculation without intermediate
negative energy states.
\end{abstract}


\section{Introduction}

The relativistic Brueckner-Hartree-Fock model \cite{Anastasio,Horowitz,terHaar,Machleidt}
is very succesful in describing equilibrium properties of nuclear matter. 
Where the conventional Brueckner-Hartree-Fock model is not able to 
reproduce the empirical saturation point without the introduction
of explicit three-body forces, the relativistic extension finds 
a saturation point very close to the experimental value.
Going beyond the relativistic Brueckner scheme and including hole-hole 
propagation to all orders, one finds only a small shift in the 
saturation point towards the empirical one \cite{FdJ_cons}. 
Including hole-hole propagation is essential for obtaining a
`conserving' model, i.e. a model that preserves conservation laws
and is thermodynamically consistent \cite{Baym}.
In this extended relativistic Brueckner model other observables
also compare favourably with experiment:
we found a compressibility in the range 250-300 MeV, a depletion 
of the Fermi-sea of $11-13 \%$ and a quasi-particle strength at 
the Fermi-surface in the order of $0.7$.

The essential ingredient in the RBHF model is the decomposition of
the self-energy in a scalar and vector part. 
While the mean-field, in essence the sum of both parts, is in the 
order of $\sim 50$ MeV, the components are a few hundred MeV. 
This leads to a small effective mass, which again results in 
a squeezing of the scalar (attractive) part of the interaction.
This additional density dependent, purely relativistic, repulsive effect 
is responsible for the shift in the saturation point as compared to 
the non-relativistic Brueckner model. 
The use of relativistic kinematics has only a minor impact. 

One problem intimately related to relativity that has still to be 
addressed is the effect of negative energy states. 
These are affected by the medium as well, and the question to what 
extent this changes the results is largely unexplored territory.
If one takes the effective Lagrangian approach on which the
RBHF model is based seriously, one should attempt to develop some 
kind of renormalization scheme and chart the effects of the medium 
on this. 
This has been done in the relatively simple framework of the Walecka 
model in the loop expansion and was only successful in the sense
that counterterms etc. can be calculated and used on a mean-field
level in Hartree and Hartree-Fock approaches on a purely 
phenomenological level \cite{Bielajew}. 
However, this loop expansion shows no signs of converging:
the coupling constants change with every additonal order 
included \cite{Furnstahl}.
In a way this failure can be expected because the energy scales involved
in any renormalization scheme will be in the order of a few GeV.
On this scale one starts to be sensitive to the substructure of 
the nucleon and therefore to the way how one suppresses this by e.g. 
a formfactor. 
Consequently, one should not try to renormalize models that are 
based on point-like nucleons.

In this work we take a less fundamental approach and avoid
the renormalization problem by using an effective theory in the 
vacuum. 
We employ a model that includes propagation of negative energy
states in a well-defined manner.
This is implemented by using the so-called spectator 
equation \cite{Gross_red}.
In this particular three-dimensional reduction of the full 
Bethe-Salpeter equation, negative-energy propagation is explicitly
included. 
In Ref. \cite{Gross} an interaction is presented that fits the 
nucleon-nucleon data well using this equation. 
By fitting the data in vacuum one renormalizes the theory 'effectively'.
This of course means that one loses the dynamics of the 
renormalization scheme in the medium. 
However, one still has the dynamics of the propagation of the 
intermediate negative energy state and it is interesting to 
investigate to what extent this affects the conclusions 
of the standard RBHF model. 

We want to stress though that before one can generalize these 
conclusions more work needs to be done. 
The three-dimensional reduction scheme we use in this work 
is not unique, and therefore any effect will be, at first, specific 
only to this particular reduction scheme. 
To arrive at more general statements one needs to investigate
the effects of negative-energy propagation in other
reduction schemes. 
However, before one can do this one needs models that fit 
nucleon-nucleon data including negative energy propagation.
We are aware of only two efforts in this direction.
One is the work of the Utrecht group \cite{Fleischer}, the other
the approach of Gross {\it et al.} \cite{Gross} which we use. 

This paper is organized as follows: we first review in section \ref{Theory} 
the theoretical basis of our model. 
In section \ref{Results} we will present results for various observables,
before we present some concluding remarks in the final section. 

\section{Theoretical Framework}
\label{Theory}

To avoid the complexity of solving the full four-dimensional Bethe-Salpeter 
equation one usually resorts to a three-dimensional reduction scheme. 
By approximating the propagator with a form that fixes the energy
variable while preserving the elastic cut, the four-dimensional integration
is reduced to a three-dimensional one over the spatial momenta. 
Even with the requirement that one retains covariance, there is no 
unique way to perform this reduction.
Over the years several specific schemes have been widely used, most notably
the Blankenbecler-Sugar (BbS) \cite{Blankenbecler} and Thompson \cite{Thompson}
reductions. 
In both these equations the intermediate nucleons are symmetrically off-shell
and no energy is transferred via the interaction, i.e., no mesonretardation
is present. 
Another point worth noting is that these equations, by construction, only 
propagate positive-energy intermediate states. 
Therefore, the main difference between these equations and the non-relativistic
scattering equation is the presence of relativistic kinematics, so-called
'minimal relativity'.
RBHF calculations based on both equations have been performed by many groups, 
we mention the pioneering work of Anastasio {\it et al.} and the 
calculations by Machleidt \cite{Machleidt}, Horowitz and Serot \cite{Horowitz} 
and the Groningen group \cite{terHaar,FdJ_cons}.

A reduction scheme that in some respects goes beyond this is the 
Gross-reduction, leading to the spectator equation \cite{Gross_red}. 
In this approach one particle is put on-shell, while in the propagation of the 
off-shell particle both the positive- and negative-energy contributions
are retained.
Also some mesonretardation is present.

With regard to three-dimensional reductions one should keep in mind that it
is not {\em a priori} clear that the Bethe-Salpeter equation using only
the One-Boson-Exchange (OBE) interaction as an approximation of the full 
irreducible kernel is superior to a well chosen three-dimensional reduction in 
describing the full scattering process.
For scalar theories it is proven that the spectator equation effectively 
includes the cancellation between direct and crossed box diagrams.
The spectator equation then gives a better approximation to the full
sum of ladder and crossed-ladder diagrams than the iterated OBE diagrams
of the Bethe-Salpeter equation \cite{Gross_red}.

In the spectator approach the 2-body propagator in the c.m. frame
reads
\begin{equation}
G_{12}(s, \bar{p}) = \frac{m^2}{E_p^2} \Lambda^{+}_1(\bar{p}) 
\left(
\frac{\Lambda^{+}_2(-\bar{p})}{\sqrt{s} - 2E_p + i\varepsilon} -
\frac{\Lambda^{-}_2(\bar{p})}{\sqrt{s} - i\varepsilon}
\right).
\label{prop_vac}
\end{equation}
The indices 1,2 identify the particle, in this propagator particle 1 is 
on-shell.
Also, $\bar{p}$ is the c.m. 3-momentum, $s$ the total invariant mass
and $\Lambda^{\pm}$ are the usual projection operators on positive/negative
energy states. 
It is easily verified that this propagator is covariant.
This also holds separately for both the positive and negative energy part. 
Note that we can obtain the Thompson equation by ignoring the negative energy 
part of the propagator and the meson retardation. 

In Ref. \cite{Gross} an OBE model based on the spectator-equation is presented.
Using only four mesons ($\pi, \rho, \omega$ and $\sigma$) a fit comparable
in quality to the Bonn potential \cite{Machleidt} was obtained.
Several potentials are given, differing in a) taking the pion-nucleon interaction
as purely pseudo-vector or with an admixture of pseudo-scalar and in 
b) the treatment of spurious poles in the meson propagators due to 
particular form of meson-retardation in the spectator equation.
The difference between a rather simple and a more elaborate scheme for the treatment
of the meson poles were shown to be relatively small. 
Since our calculations are numerically already very involved we opt for the
simple scheme and take the IA model of Ref. \cite{Gross} for our calculations 
in the nuclear medium.

In isotropic symmetric nuclear matter the self-energy of the nucleon has
in the spinor representation several components
\begin{equation}
\Sigma(p) = \Sigma^{\rm s}(p) - \gamma^0 \Sigma^0(p) + 
\bar{p} \cdot \bar{\gamma} \Sigma^{\rm v}(p).
\end{equation}
This decomposition of the Dirac structure allows us formally to define 
effective (in principle momentum and energy dependent) quantities
\begin{eqnarray}
m^* &=& m + \Sigma^{\rm s}(p) \nonumber\\
p_0^* &=& p_0 + \Sigma^0(p) \nonumber\\
\bar{p}^* &=& \bar{p} \left[1 + \Sigma^{\rm v}(p) \right].
\end{eqnarray}
Previous calculations have shown that the momentum dependence of the 
self-energy components is relatively weak. 
We follow the standard procedure to neglect the momentum 
dependence of the self-energies in the intermediate propagators
by using the values at $p = p_f$.
Furthermore, from a technical viewpoint it is advantageous to 
`divide out' the $\Sigma^{\rm v}$ component. 
One then obtains a renormalized effective mass:
\begin{equation}
m^*_{\rm sc} = \frac{m + \Sigma^{\rm s}(p_f)}{1 + \Sigma^{\rm v}(p_f)},
\end{equation}
and corresponding expressions are obtained for momentum and 
vector self-energies.
This effective mass will be used in the practical calculation, where 
confusion might arise we will refer to this particular definition
as the self-consistent effective mass.
With this definition we find an effective positive-energy spinor:
\begin{equation}
u^*_r(\bar{p}) = {\left(\frac{E^*_p + m^*}{2m^*}\right)}^{\frac{1}{2}}
{
\left(
 \begin{array}{c}
 1 \\
 \displaystyle{{\frac{\bar{\sigma}\cdot{\bar{p}}}{E^*_p + m^*}}}
 \end{array}
\right)}
\chi_r.
\label{effspinor}
\end{equation}
Here $E^* = \sqrt{\bar{p}^2 + {m^*}^2}$ is the on-shell energy of the
spinor.
As we will see later the value of $\Sigma^{\rm v}$ is still small, but not
negligible as in earlier calculations and we have to incorporate it in 
the calculation. 
One thus finds for the dressed one-body propagator \cite{Amorim} (note
that in this expression $p_0^*$ is also renormalized by 
$1 + \Sigma^{\rm v}(p_f)$)
\begin{equation}
g(p) = 
\frac{1}{1 + \Sigma^{\rm v}}\frac{m^*}{E^*_p}
\left[
\frac{\Lambda^{*+}(p) \Theta(p - p_f)}{p_0^* - E^*_p + i \varepsilon} +
\frac{\Lambda^{*+}(p) \Theta(p_f - p)}{p_0^* - E^*_p - i \varepsilon} -
\frac{\Lambda^{*-}(-p)}{p_0^* + E^*_p - i \varepsilon}
\right].
\label{prop1_med}
\end{equation}
Where $\Lambda^{*+}$ is the projector on effective positive-energy
states: $u^*(p) \otimes \bar{u}^*(p)$, with the effective spinors of
Eq. (\ref{effspinor}). 
In line with the Brueckner scheme we neglect the hole ($p < p_f$) part
in the two-body propagator in the medium. 
We find \cite{Amorim}
\begin{equation}
G_{12}(\sqrt{s^*}, P, \bar{p}) = \frac{{m^*}^2}{(1 + \Sigma^{\rm v}){E_p^*}^2} 
\Lambda^{+*}_1(\bar{p}) 
\left(
\frac{Q_{pp}(P, \sqrt{s^*}, p) \Lambda^{*+}_2(-\bar{p})}{\sqrt{s^*} - 2E^*_p + i\varepsilon} -
\frac{Q_{p}(P, \sqrt{s^*}, p) \Lambda^{*-}_2(\bar{p})}{\sqrt{s^*} - i\varepsilon}
\right).
\label{prop2_med}
\end{equation}
In this equation $Q_{pp}(P, \sqrt{s^*}, p), Q_p(P, \sqrt{s^*}, p)$ are the relativistic 
angle-average Pauli-blocking operators \cite{terHaar}, respectively blocking 
out two and one momenta below the Fermi-momentum.
They depend on the total invariant effective mass $\sqrt{s^*}$, the
total momentum $P$ and the blocked momentum $p$.
Both momenta in the particle-particle part of the propagator are required
to have momenta above $p_f$, in the particle-antiparticle part only the
momentum of the (positive energy) particle is Pauli blocked. 

Any incoming, outgoing or intermediate state can either be 
particle-particle (denoted by '+') or particle-antiparticle (denoted by '-')
character. 
Splitting the two-body propagator in two contributions using this nomenclature we
find a set of coupled equations for the various components of the 
T-matrix:
\begin{eqnarray}
T^{++}_{\rm dir} &=& V^{++}_{\rm dir} + 
V^{++}_{\rm dir} g_{1}^+ T^{++}_{\rm dir} + V^{++}_{\rm exch} g_{2}^+ T^{++}_{\rm exch} +
V^{+-}_{\rm dir} g_{1}^- T^{-+}_{\rm dir} + V^{+-}_{\rm exch} g_{2}^- T^{-+}_{\rm exch},
\nonumber\\
T^{-+}_{\rm dir} &=& V^{-+}_{\rm dir} + 
V^{-+}_{\rm dir} g_{1}^+ T^{++}_{\rm dir} + V^{-+}_{\rm exch} g_{2}^+ T^{++}_{\rm exch} +
V^{--}_{\rm dir} g_{1}^- T^{-+}_{\rm dir} + V^{--}_{\rm exch} g_{2}^- T^{-+}_{\rm exch}.
\end{eqnarray}
The terminology {\em dir/exch} arises from which particle is on-shell 
(see also Ref. \cite{Gross}): $V^{\pm \pm}_{\rm dir}$ is an interaction where 
both on-shell particles are on the same-vertex, in $V^{\pm \pm}_{\rm exch}$
each vertex has one on-shell particle. 
In the equation above the outgoing upper particle is always on-shell.
Which particle is on-shell in the intermediate state is indicated by the index 
in the two-body propagator.
The expressions for $T^{\pm +}_{\rm exch}$ have a similar structure. 
In Ref. \cite{Gross} it is shown that the resulting scattering amplitude is 
properly anti-symmetrized.
By using the form of the positive/negative energy projection 
operators ($\Lambda^{*+} = u^*(p) \otimes \bar{u}^*(p)$) the above set of
equations can be expressed in momentum spin-space. 
In the c.m. frame we then have, e.g.,
$V^{++}_{\rm dir} = \bar{u}_3^*(p')\bar{u}_4^*(-p') V u_1^*(p)u_2^*(-p)$
and 
$T^{+-}_{\rm dir} = \bar{u}_3^*(p')\bar{v}_4^*(-p') T u_1^*(p)u_2^*(-p)$
and so on.
The T-matrix is calculated in this representation.
We like to stress again that $V$ is sandwiched between {\it effective} spinors.
Due to the low effective nucleon mass
this leads to a supression of the attractive scalar part and an
enhancement of the repulsive vector part of the interaction.
This addional effective (density dependent) repulsion is a 
consequence of the relativistic treatment and not present 
in non-relativistic Brueckner models, 
it is the reason for the empirical succes of the 
relativistic Brueckner models.

Due to the kinematics introduced by the specific form of the spectator 
equation, $V^{\pm \pm}_{\rm exch}$ can develop a spurious meson-production
pole. 
To counter this we use scheme I of Ref. \cite{Gross}.
It amounts to replacing the 'exchange' kinematics by 'direct' kinematics
in the region where the spurious pole can occur.
This can be achieved preserving the continuity of $V^{\pm \pm}_{\rm exch}$.

With the $T$-matrix we calculate the self-energy in the Bruekner
approximation:
\begin{equation}
\Sigma = \int \frac{d^4p}{(2 \pi)^4}
Re \left[ T^+_{\cal A} \right] g^<(p).
\label{def_sig}
\end{equation}
With $T^+_{\cal A}$ the antisymmetrized $T$-matrix in spinor-space, 
the `$+$' indicates that we only take the part which is analytic in the upper 
complex $\sqrt{s}$ plane.
In order to obtain the Brueckner contribution to the self-energy we take 
for the correlation propagator $g^<(p)$ only the positive energy part
\begin{equation}
g^<(p) = 2 \pi \frac{m^*}{E^*_p} \Lambda^{*+}(\bar{p}) \delta(p_0^* - E_p^*).
\end{equation}
By using this expression for the self-energy we eliminate contributions from 
the Dirac-sea and avoid the problems due to the renormalization of these.

As is clear from Eq. (\ref{prop2_med}) the full $T$-matrix has poles in the upper
and lower complex $\sqrt{s^*}$ plane.
The poles in the upper halfplane are generated by the particle-antiparticle 
propagating part, the poles in the lower halfplane by the particle-particle
propagating part. 
These poles are well separated as a function of $\sqrt{s^*}$ and we can split
the imaginary part of $T$ in parts with poles only in the upper or lower half
plane
\begin{equation}
Im (T) = Im (T^+) + Im (T^-),
\end{equation}
where $T^{+,-}$ are analytic in the upper and lower half-plane respectively. 
This allows us to apply Cauchy's theorem for both contributions separately by 
either closing the contour over the upper or lower halfplane.
From Eq. (\ref{prop2_med}) it follows that $Im (T^-)$ vanishes on the
real axis, apart from the point $\sqrt{s^*} = 0$. 
This is a highly singular point, 
we avoid it by using the contour depicted in Fig. \ref{Figure_contour}.
Straightforward application of Cauchy's theorem gives for real $\sqrt{s^*}$
\begin{equation}
Re \left[ T^-(\sqrt{s^*}) \right] = - \frac{1}{\pi} 
\int_C Im \left( \frac{T^-(\xi) \xi'}{\xi - \sqrt{s^*}} \right).
\label{calc_tmin}
\end{equation}
The integrand is zero everywhere except on the halfarc of the contour.
There we have to calculate the $T$-matrix with complex values of $\sqrt{s^*}$ in the
propagator. 
We now calculate $Re(T^+)$ by
\begin{equation}
Re(T^+) = Re(T) - Re(T^-).
\label{def_tplus}
\end{equation}
We tested this method by setting the particle-particle propagating part of 
the 2-body propagator of Eq. (\ref{prop2_med}) equal to zero. 
Then $Re(T^+)$ should be equal to the Born contribution. 
As it turns out, we need only 6 integration points on the arc to obtain
very good accuracy. 
Another test is setting the particle-antiparticle propagating part equal to
zero, than $Re(T^-)$ calculated via Eq. (\ref{calc_tmin}) should vanish as 
well, which is again very well fulfilled.

In Eq. (\ref{def_sig}) the $T$-matrix is needed in spinor representation.
However, in the practical calculation we have the $T$-matrix in spin
representation. 
To extract the spinor representation of the self-energy several (approximate) 
methods have been used in the past.
Assuming momentum-independent self-energies one can extract the self-energy 
components by the momentum dependence of the single particle energy
\cite{Machleidt}. 
Another possibility is to project the $T$-matrix elements in 
the positive-energy sector of the spin space on 
Lorentz invariant amplitudes as is done e.g. in 
Refs. \cite{Horowitz,terHaar,FdJ_cons}.
Both methods are not exact, for the projection method to be unique one 
needs to include also the negative-energy sector of the spin 
space \cite{Amorim}.
In the present calculation we want to decompose the self-energy
unambiguously. 
This is achieved by calculating the full set of quadratic forms
$\bar{u}^*(\bar{p}) \Sigma(\bar{p}) u^*(\bar{p})$,
$\bar{u}^*(\bar{p}) \Sigma(\bar{p}) v^*(\bar{p}) = $
$\bar{v}^*(\bar{p}) \Sigma(\bar{p}) u^*(\bar{p})$ and
$\bar{v}^*(\bar{p}) \Sigma(\bar{p}) v^*(\bar{p})$.
Working out the straightforward Dirac algebra
in a similar fashion as in Refs. \cite{Anastasio,Poschenrieder} we find
\begin{eqnarray}
\Sigma^{\rm s}(\bar{p}) &=& \frac{1}{2} 
\left[
\bar{u}^*(\bar{p}) \Sigma(\bar{p}) u^*(\bar{p}) +
\bar{v}^*(\bar{p}) \Sigma(\bar{p}) v^*(\bar{p})
\right]
\nonumber\\
\Sigma^{0}(\bar{p}) &=& -\frac{E^*_p}{2m^*}
\left[
\bar{u}^*(\bar{p}) \Sigma(\bar{p}) u^*(\bar{p}) -
\bar{v}^*(\bar{p}) \Sigma(\bar{p}) v^*(\bar{p})
\right]
+ \frac{p}{m^*} \bar{v}^*(\bar{p}) \Sigma(\bar{p}) u^*(\bar{p}),
\nonumber\\
\Sigma^{\rm v}(\bar{p}) &=& - \frac{1}{2m^*} 
\left[
\bar{u}^*(\bar{p}) \Sigma(\bar{p}) u^*(\bar{p}) -
\bar{v}^*(\bar{p}) \Sigma(\bar{p}) v^*(\bar{p})
\right]
+ \frac{E^*_p}{m^*p} \bar{v}^*(\bar{p}) \Sigma(\bar{p}) u^*(\bar{p}).
\end{eqnarray}
Finally, after working out the integration in Eq. (\ref{def_sig})
and transforming to c.m. defining variables ($s^*,P$)
we have to calculate
(note that $T^{+}(s^*,P)$ transforms covariantly)
\begin{equation}
\bar{w}^*_r(\bar{p}) \Sigma(\bar{p}) y_r^*(\bar{p})
=
\int_{P^2_{\rm min}}^{P^2_{\rm max}} dP^2
 \int_{s^*_{\rm min}}^{s^*_{\rm max}} ds^* 
 \frac{1}{16 \pi^2 p} \frac{m^*}{E^*_p + E^*_q}
{\rm Tr}_{s} 
\left[
\bar{w}^*_r(\bar{p}) \bar{u}^*_s(\bar{q}) T^{+}_{\cal A}(s^*,P)
y_r^*(\bar{p}) u_s^*(\bar{q})
\right].
\end{equation}
In this equation $w_r$ and $y_r$ can be either a 
positive or negative-energy spinor, $r$ is the spin-index,
$q$ is the momentum of the integrated particle in the 
nuclear matter rest-frame, found by
applying the inverse transform to the set ($s^*,P$).
The integration limits are given by
$P_{\rm max} = p + p_f$, $P_{\rm min} = \max(0, p - p_f)$,
$s^*_{\rm min} = (E^*_p + E^*_{q_{\rm min}})^2 - P^2$ with 
$q_{\rm min} = |P - p|$ and $s^*_{\rm max} = 4{E^*_p}^2 - P^2$.
The trace in the integrand is over the spin-index $s$ of the 
integrated particle.

Before we turn to the results, one remaining point of the model has
to be addressed. 
In the OBE interaction of Ref. \cite{Gross} on top of the 
common mesonic formfactors also a nucleonic formfactor, depending
on the off-shellness of the nucleon, is introduced. 
It suppresses large off-shell energies and momenta which probe 
sub-nucleonic structures and reads 
\begin{equation}
h(p^2) = \frac{2(\Lambda_n^2 - m_n^2)}{(\Lambda_n^2 - p^2) + 
(\Lambda_n^2 - m_n^2)}.
\end{equation}
The question is how to treat this in the medium. 
Since the formfactor lacks a microscopic foundation (apart from 
the fact that it can be related to an energy-dependent self-energy),
we have little guidance on how it changes in the medium. 
Since the on-shell relation in the medium reads
${p^*}^2 = {m^*}^2$ it seems appropriate to use
\begin{equation}
h(p^2) = \frac{2({\Lambda_n^*}^2 - {m_n^*}^2)}
{({\Lambda_n^*}^2 - {p^*}^2) + ({\Lambda_n^*}^2 - {m_n^*}^2)},
\end{equation}
where we scale $\Lambda_n^*$ with the effective mass. 
This idea is slightly generalized by using 
\begin{equation}
\Lambda_n^* = \frac{\alpha (m^*_n - m_n)  + m_n}{m_n} \Lambda_n.
\end{equation} 
Taking $\alpha$ equal to zero leaves $\Lambda_n^*$ equal to the vacuum 
value, $\alpha$ = 1 fully rescales $\Lambda_n^*$ with the nucleon effective 
mass.

This completes the theoretical framework of our model. Given 
a density we now can iterate around the self-energy 
evaluated at the Fermi-surface until a self-consistent 
solution is found. 
With these self-consistent values we can calculate relevant observables.

\section{Results}
\label{Results}

In this section we present results for our model with explicit intermediate
negative energy states. 
We compare our results with those obtained using the Thompson reduced 
version of the Bonn-C potential \cite{Machleidt}. 
The latter model only contains positive-energy intermediate states.
However, both models do fit the nucleon-nucleon scattering data 
in the vacuum equally well. 
As we will see, the differences in the underlying dynamical description
will lead to differences in the results.
Of course, the potentials of the models are different as well, but 
this is a consequence of the different dynamics from the requirement
that one fits the nucleon-nucleon scattering data.

We first calculate the equation of state, the results are presented in 
Fig. \ref{Figure_Eb}.
The full line obtained with the cutoff-renormalization parameter 
$\alpha = 1$, the dashed line with $\alpha = 0.95$. 
We see that the dependence on this parameter is (unfortunately) rather 
large. 
Moreover, lowering $\alpha$, and thereby letting $\Lambda_n$ approach its
vacuum value, makes the system more attractive. 
If the system gets too attractive, which happens at normal nuclear density 
for $\alpha \sim 0.8$, it develops an instability which precludes 
a calculation of the T-matrix.
As a reference we also plot the result obtained with the 
Thompson-reduction version of the Bonn-C potential \cite{Machleidt}.
In this calculation we used the uniquely determined self-energies and 
did include the $(1 + \Sigma^{\rm v})$ factor in the 2-body propagator
cf. Eq. (\ref{prop2_med}).
One observes that the result from the spectator equation is much stiffer 
than the Bonn-C result. 
This is partly expressed by the compressibility at saturation, 180 MeV for
the Bonn-C result and 250 MeV for the $\alpha = 0.95$ result.
However, one should be careful with comparing compressibilities calculated
for different saturation points. 
The figure does seem to indicate that for higher densities the equation of
state of the spectator equation is stiffer. 
A probable explanation for this is due to the low effective mass.
The particle-antiparticle propagating part provides a large part of 
the repulsion \cite{Gross}. 
With a lower effective mass, the incoming $\sqrt{s^*}$ becomes smaller as well and
the $T$-matrix equation scales in a sense: the pole of the antiparticle-particle
part of the propagator comes closer to the pole of the particle-particle part.
This leads to a larger influence of the particle-antiparticle propagating part 
and therefore to more repulsion. 
This is an additional repulsive effect of the low effective nucleon mass,
on top of the conventional relativistic effect of squeezing the attractive 
scalar part of the interaction. 

We want to stress that this additional repulsion is a direct consequence of 
the particular three-dimensional reduction we use. 
As pointed out in Ref. \cite{Gross} in fitting the nucleon-nucleon scattering 
data the dynamical repulsion of the particle-antiparticle intermediate 
states results in a relatively low coupling constant of the (repulsive) 
$\omega$-meson as compared to other potentials.
Now, only part of the total repulsion needed to fit the data has to 
come from the $\omega$-meson.
In the Bonn-C on has $g^2_\omega/4\pi = 20$ while in our OBE we use
$g^2_\omega/4\pi = 9.85$.
The `difference' between these values is now included dynamically in the 
model, and the medium effects on this result in differences as we 
pointed out above. 
Note that a lower $g_\omega$ brings it in better agreement with the
values obtained in pion-nucleon photoproduction models, see e.g. Ref.
\cite{Nozawa,Davidson}.

In Fig. \ref{Figure_dens_dep} we show the density dependence of the effective
mass and self-energy components. Our findings are in line with conventional 
Dirac-Brueckner results. 
The value for the effective mass is similar to what is found in other calculations,
we also find the usual linear dependence on the Fermi-momentum. 
The same is valid for $\Sigma^{\rm s}$ and $\Sigma^0$. 
The values for these are a bit smaller than is usually found, this is mainly due
to the fact that we determine the self-energy components uniquely. 
We also find that the isospin $I= 1$ channel contributes the most to the
respective components of the self-energy. 
At normal nuclear matter density $\Sigma^{\rm s} = -237$ MeV, 
$\Sigma^{0} = -197$ MeV in the $I= 1$ channel, while in the $I= 0$ channel
the respective values are $\Sigma^{\rm s} = -49$ MeV and 
$\Sigma^{0} = -10$ MeV, 
although the contribution of the channels to the mean-field 
($\bar{u}^* \Sigma u^*$) at the Fermi-surface is comparable: 
$-25$ MeV and $-6$ for the $I = 0,1$ channels respectively.
In an earlier calculation we found the same when using the projection
method in extracting the self-energies \cite{terHaar,FdJ_cons}.
A notable difference with other calculations is the value of 
$\Sigma^{\rm v}$: we find a positive and rather large value. 
The positive value is a consequence of the proper, unique, decomposition of
the self-energy, similar results are found in Refs. \cite{Poschenrieder,Amorim}
and also in our recalculation of the Bonn-C results.
The absolute size of $\Sigma^{\rm v}$ is a particular feature of our 
calculation.
It makes the inclusion of the $1 + \Sigma^{\rm v}$ factor in the 2-body
propagator Eq. (\ref{prop2_med}) necessary. 

It is worthwhile to note that the use of a non-unique method of extracting 
the self-energy components like e.g. the projection method of Refs. 
\cite{Horowitz,terHaar} and the momentum-dependence method of Ref. 
\cite{Machleidt} leads to differences which, 
up to order ${\cal O}(p^4/{m^*}^4)$, only depend on one parameter.
Two different decompositions of the self-energy $\Sigma,\Sigma'$
need to give the same value for the mean-field: 
$\bar{u}^*(\bar{p}) \Sigma(\bar{p}) u^*(\bar{p}) = 
\bar{u}^*(\bar{p}) \Sigma'(\bar{p}) u^*(\bar{p})$.
From this is follows that up to ${\cal O}(p^4/{m^*}^4)$ 
$\Sigma_s' = \Sigma_s - \alpha$,  
$\Sigma_0' = \Sigma_0 - \alpha$ and  
$\Sigma_v' = \Sigma_v - \alpha/2m^*$.
The `ambiguity' coefficient $\alpha$ depends of course on density and 
momentum.
For example, the momentum-dependence method sets $\Sigma_v' = 0$, which 
implies that $\alpha = 2m^* \Sigma_v$, with $\Sigma_v$ the exact
component.
The projection method tends to give negative values for $\Sigma_v'$, whereas
the exact value is positive.
This gives an even larger value for the `error' coefficient $\alpha$. 
Our unique decomposition now allows us to calculate the 
mean-field effective coupling constants in a more proper way than 
we did in Ref. \cite{FdJ_asym}. 
In Fig. \ref{Figure_eff_coupl} we show the effective isoscalar scalar 
and vector coupling constants.
These are defined by
\begin{equation}
\left(\frac{g^*_{\sigma}}{m_{\sigma}}\right)^2 
= -\frac{1}{2} 
\frac{\Sigma^{s}_{n}(\bar{p}_f) + \Sigma^{s}_{p}(\bar{p}_f)}{\rho^s},
\left(\frac{g^*_{\omega}}{m_{\omega}}\right)^2 
= -\frac{1}{2} 
\frac{\Sigma^{0}_{n}(\bar{p}_f) + \Sigma^{0}_{p}(\bar{p}_f)}{\rho^v}.
\end{equation}
where $\rho^s$ and $\rho^v$ are the usual scalar and 
vector densities from relativistic mean-field theory \cite{SW}.
Comparing the Bonn-C result with those obtained in Ref. \cite{FdJ_asym}
using the projection method we find that the exact values are smaller 
and show less dependence on the density. 
The effective scalar coupling only decreases slightly as a function of density,
the effective vector coupling even increases slightly with 
density.
Both effective couplings are still linear in the Fermi-momentum.
The appreciable differences between the projection method and the 
exact results show that for detailed results one can not rely on an 
approximative method and has to calculate the unique decomposition.
The result of the present calculation with the spectator equation shows
a much stronger density dependence, also linear in the Fermi-momentum.
The most striking feature is that both couplings increase with 
increasing density
This again reflects the additional dynamics due to negative
energy intermediate states.

The momentum dependence of the self-energy components is shown in Fig. 
\ref{Figure_mom_dep} for normal nuclear matter density, corresponding 
to $p_f = 0.27$ GeV/c.
In this calculation we take $\alpha = 1$,
note that we now present $p \Sigma^v(p)$.
As a reference we give in the upper panel the results for the Bonn-C potential
and in the lower panel the results of the present work with the spectator 
equation. 
For the Bonn-C results the conventional assumption of the weak momentum 
dependence of the self-energy, which we used as well, seems to be justified,
at least for momentum within and slightly above the Fermi-sea. 
For example, $\Sigma^{\rm s}$ increases from $-320$ MeV for momenta deep in the
Fermi-sea to $-297$ MeV for $p = p_f$ and for the highest momentum give 
we find $-234$ MeV. 
The results obtained with the spectator equation show a more pronounced 
momentum dependence, up to a point where one might start to question the
usual assumption of weak momentum-dependence of the self-energy components.
We get for small momenta $\Sigma^{\rm s} = -344$ MeV, at the Fermi surface
we have $-286$ MeV, while for the largest momentum we obtain $-186$ MeV. 
Again we see that, compared with the Bonn-C results, the contribution 
from $\Sigma^{\rm v}$ is significantly larger for the present results obtained
with the spectator equation.

\section{Conclusions}

We presented a relativistic Brueckner calculation using a specific 
three-dimensional reduction of the Bethe-Salpeter equation that includes
explicit negative energy-states in the intermediate 2-body propagator. 
We employ a potential that fits the nucleon-nucleon data well using this 
equation as presented in Ref. \cite{Gross}.
The additional dynamics of the included negative-energy states lead to 
an effective repulsion, expressed by a relatively low 
$\omega$-meson coupling constant. 
To obtain the Brueckner contribution to the self-energy we need to 
separate out the part of the $T$-matrix which is analytic in the 
upper half-plane, i.e. the part generated by propagation of 
particle-particle states.
In our calculation we uniquely decompose the self-energy in 
its Dirac structure, removing the ambiguity in determining the
components present in other calculations \cite{Horowitz,terHaar,FdJ_cons,Machleidt}. 
We find a binding energy curve that compares well qualitatively 
with previous calculations.
This curve does depend significantly on how we treat the 
off-shell form-factor of the nucleon.
There are, however, differences in important details. 
Our equation of state if stiffer than found in other Dirac-Brueckner calculations. 
We argued that this is probably an additional effect of the
low effective mass. 
This results in the particle-antiparticle propagation, which
is repulsive, to become more important.
The extended dynamics introduced by the particle-antiparticle propagation,
which is a direct consequence of the particular 3-dimensional 
reduction of the full Bethe-Salpeter equation we use, also expresses
itself in the medium.
Also, and probably related to this, we find the momentum dependence
of the self-energy to be relatively large. 
It is nearly twice of what we find when using the 
Thompson-reduction version of the Bonn-C potential. 

One troubling feature of the model is the medium effect of the
off-shell nucleon formfactor. 
There is no microscopic description of this formfactor, and hence
we do not know its medium dependence. 
We used a simple renormalization of the cutoff and found 
reasonable results. 
However, the results do depend on the way how we treat
the cutoff in the medium. 
This reflects the fact that in our effective theory we are sensitive on 
how we suppress energies that probe sub-nuclear structures. 
We can only fix this at one density, taken to be the vacuum by 
reproducing the nucleon-nucleon scattering data. 
However, one then loses the dynamical description of the procedure in 
the medium.  
This question needs to be addressed in future studies.
\\
\\
\noindent
Supported by BMBF, DFG (contract Le 439/4-1) and GSI-Darmstadt

\begin{figure}
\caption{
The contour used to extract the value of $Re(T^-)$ as explained 
in the text.
}
\label{Figure_contour}
\end{figure}


\begin{figure}
\caption{
Binding energies for various values of the rescaling parameter
$\alpha$: the solid line is the result with $\alpha = 1$, the dashed 
line is calculated with $\alpha = 0.95$. 
For reference the result for the Bonn-C (Thompson reduction) using the
unique decomposition of the self-energy is given by the dotted line.
}
\label{Figure_Eb}
\end{figure}


\begin{figure}
\caption{
Density dependence of the self-consistent effective mass (solid line),
$\Sigma^{\rm s}(p_f)$ (dashed line), $\Sigma^0(p_f)$ (dotted line)
and $\Sigma^{\rm v}(p_f)$ (dash-dotted line). Note the 
relatively large and positive value of $\Sigma^{\rm v}(p_f)$.
}
\label{Figure_dens_dep}
\end{figure}


\begin{figure}
\caption{
The effective isoscalar scalar and vector coupling constants as
defined in the text. The solid line is the result obtained with the
spectator equation, the dashed line is found using the Thompson version
of the Bonn-C potential. 
}
\label{Figure_eff_coupl}
\end{figure}


\begin{figure}
\vspace*{10mm}
\caption{
Momentum dependence of the self-energy calculated at normal nuclear 
matter density, corresponding to $p_f = 0.27 GeV/c$.
In the upper panel the results for the Bonn-C potential are shown, 
in the lower panel the results for the spectator equation with 
negative energy intermediate states.
In both panels the solid line stands for $\Sigma^{\rm s}(p)$, the 
dashed line for $\Sigma^{0}(p)$ and the dotted line for 
$p\Sigma^{\rm v}(p)$.
}
\label{Figure_mom_dep}
\end{figure}

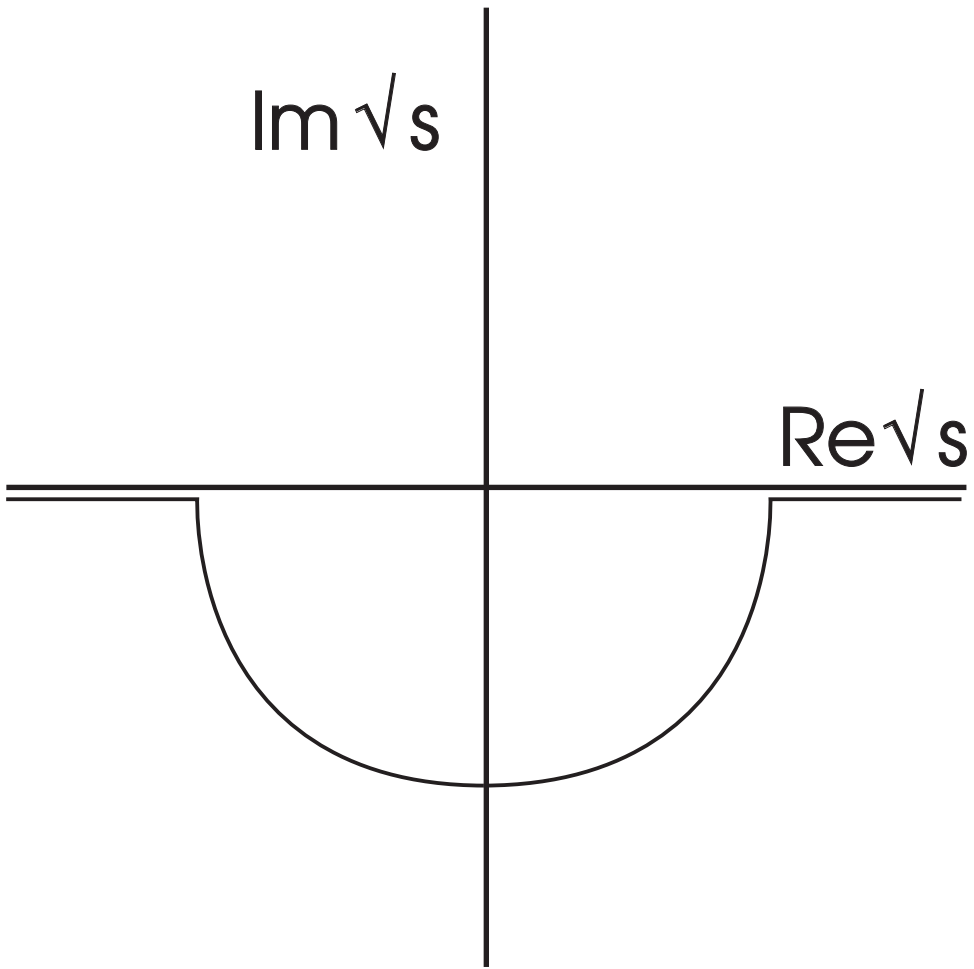
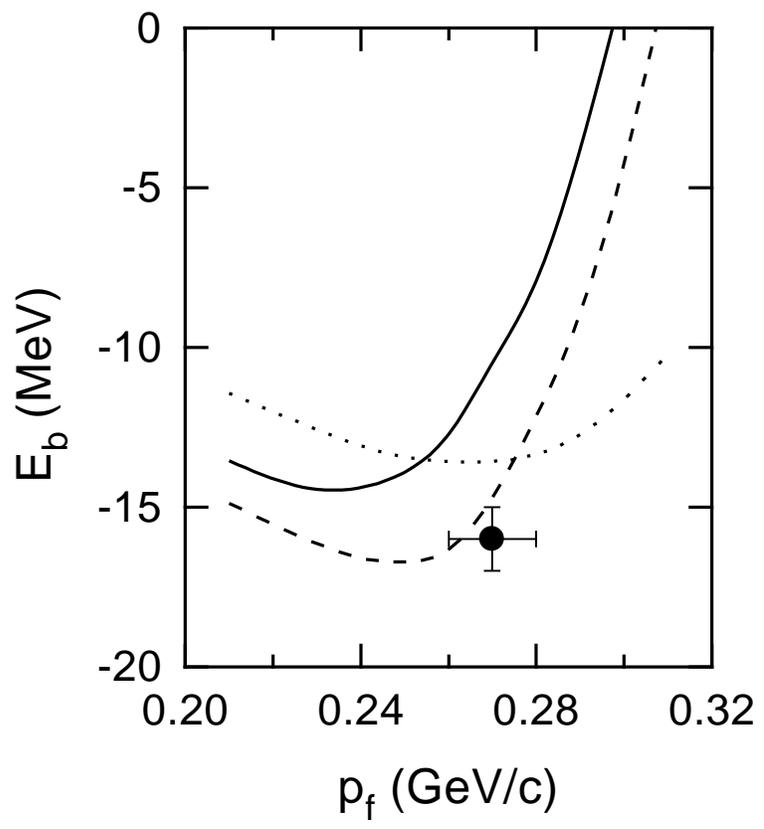
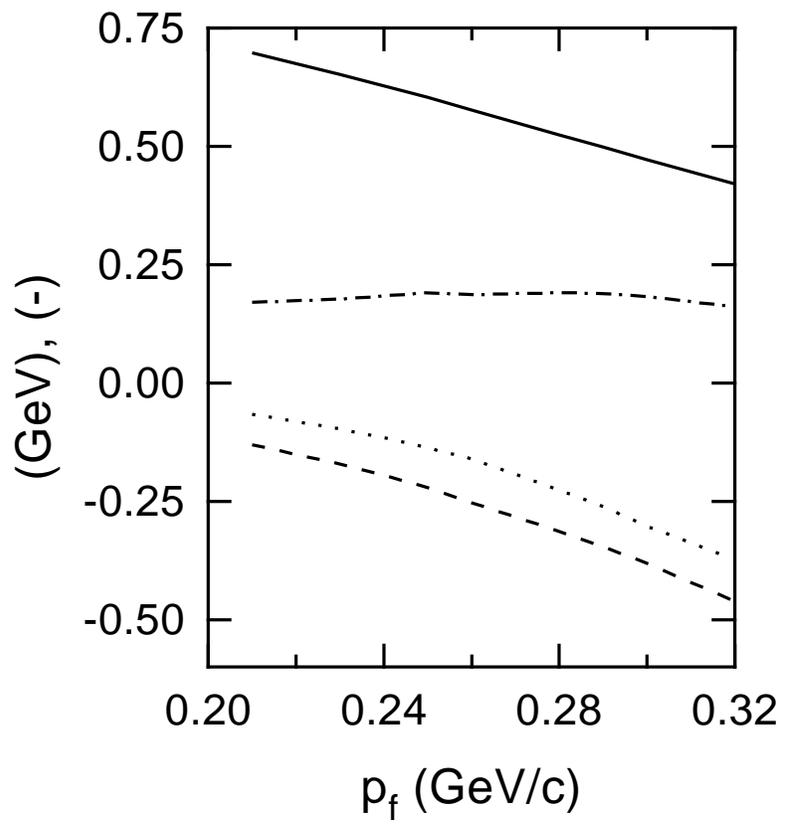
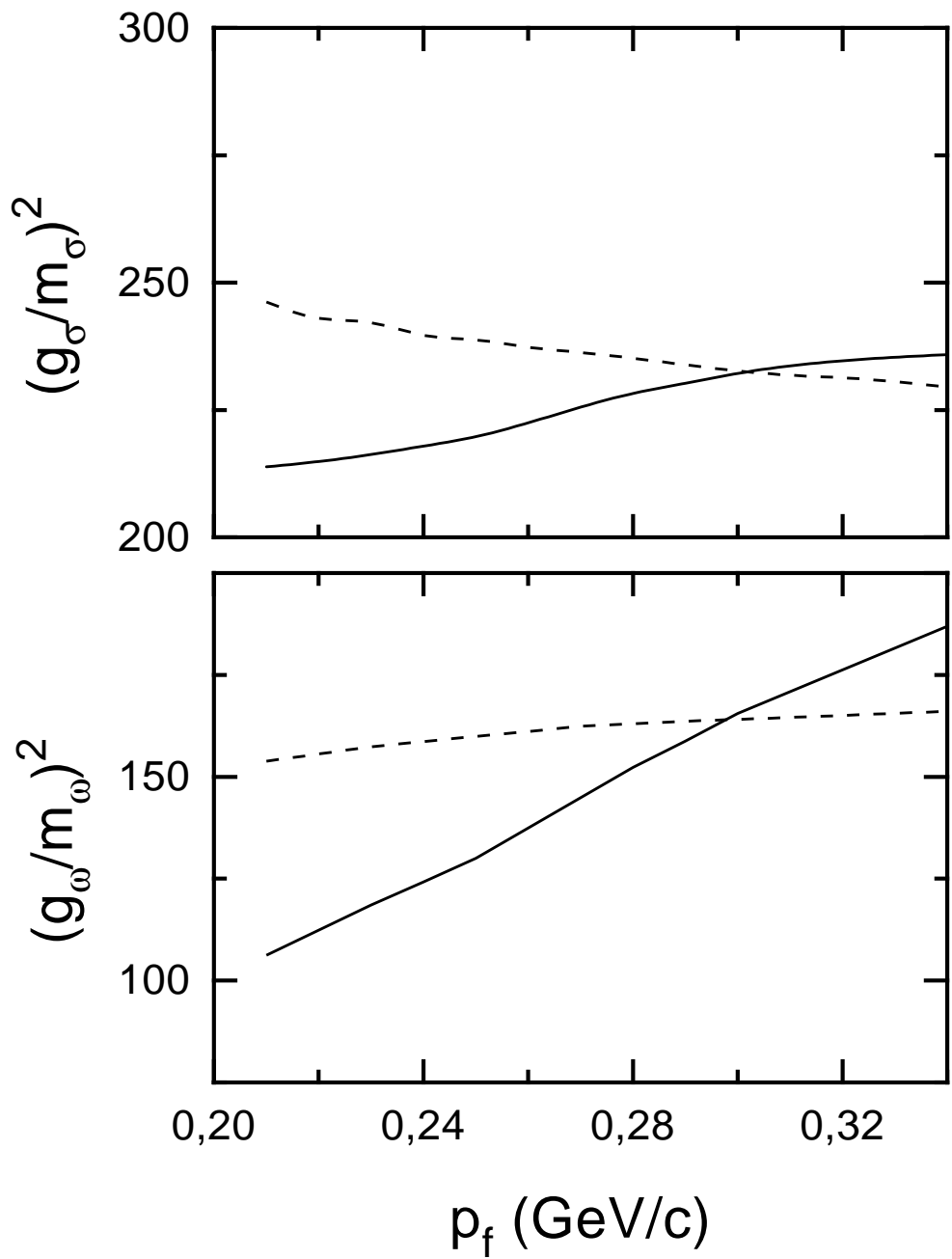
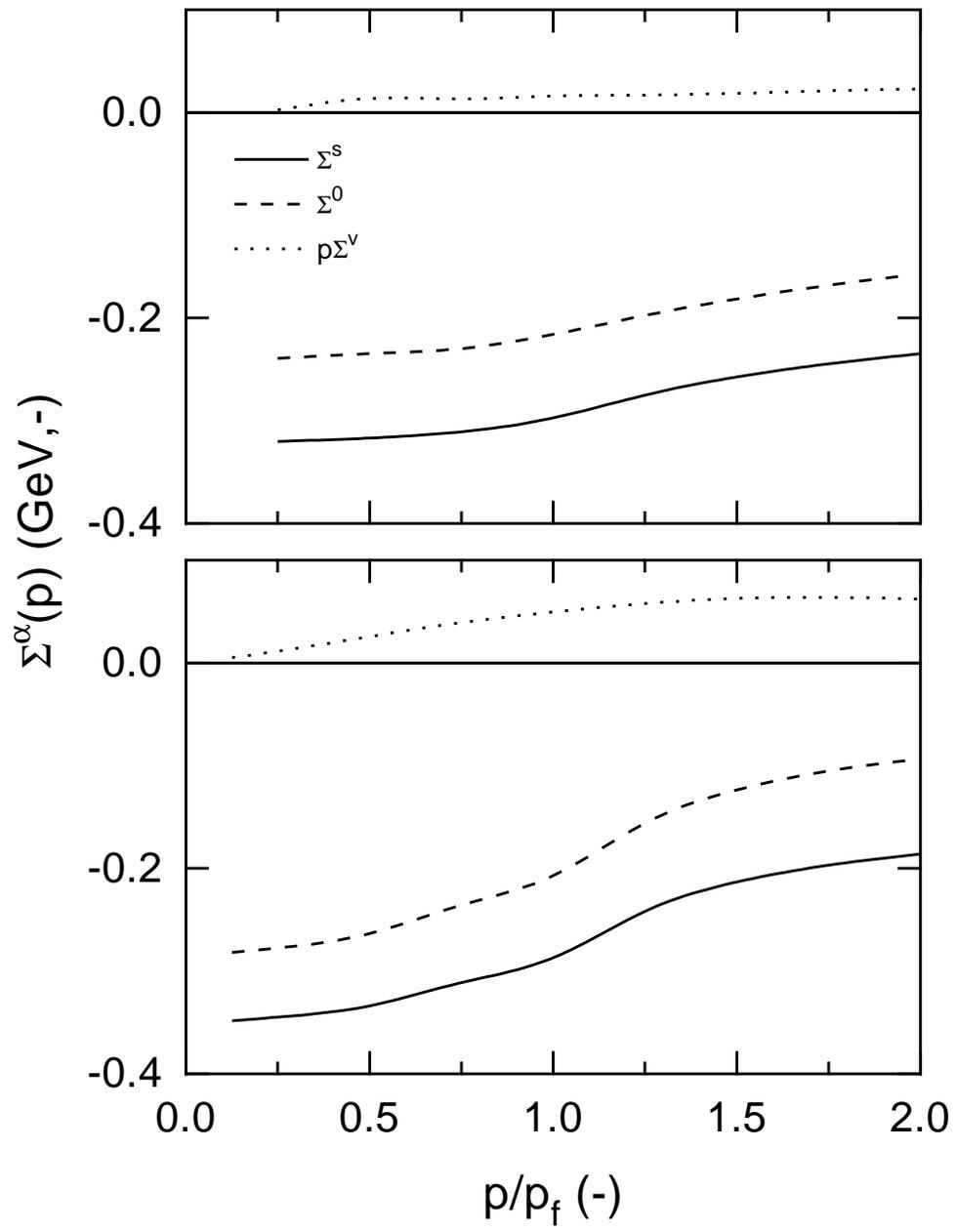
\end{document}